\documentstyle[twoside]{article}

\catcode`\@=11
\long\def\@makefntext#1{
\protect\noindent \hbox to 3.2pt {\hskip-.9pt
$^{{\eightrm\@thefnmark}}$\hfil}#1\hfill}               

\def\@makefnmark{\hbox to 0pt{$^{\@thefnmark}$\hss}}    

\def\ps@myheadings{\let\@mkboth\@gobbletwo
\def\@oddhead{\hbox{}
\rightmark\hfil\eightrm\thepage}
\def\@oddfoot{}\def\@evenhead{\eightrm\thepage\hfil
\leftmark\hbox{}}\def\@evenfoot{}
\def\sectionmark##1{}\def\subsectionmark##1{}}

\oddsidemargin=\evensidemargin
\newcounter{sectionc}\newcounter{subsectionc}\newcounter{subsubsectionc}
\renewcommand{\section}[1] {\vspace{12pt}\addtocounter{sectionc}{1}
\setcounter{subsectionc}{0}\setcounter{subsubsectionc}{0}\noindent
        {\tenbf\thesectionc. #1}\par\vspace{5pt}}
\renewcommand{\subsection}[1] {\vspace{12pt}\addtocounter{subsectionc}{1}
      \setcounter{subsubsectionc}{0}\noindent
      {\bf\thesectionc.\thesubsectionc.{\kern1pt \bfit #1}}\par\vspace{5pt}}
\renewcommand{\subsubsection}[1]
      {\vspace{12pt}\addtocounter{subsubsectionc}{1}
      \noindent{\tenrm\thesectionc.\thesubsectionc.\thesubsubsectionc.
      {\kern1pt \tenit #1}}\par\vspace{5pt}}
\newcommand{\nonumsection}[1] {\vspace{12pt}\noindent{\tenbf #1}
        \par\vspace{5pt}}

\newcounter{appendixc}
\newcounter{subappendixc}[appendixc]
\newcounter{subsubappendixc}[subappendixc]
\renewcommand{\thesubappendixc}{\Alph{appendixc}.\arabic{subappendixc}}
\renewcommand{\thesubsubappendixc}
        {\Alph{appendixc}.\arabic{subappendixc}.\arabic{subsubappendixc}}

\renewcommand{\appendix}[1] {\vspace{12pt}
        \refstepcounter{appendixc}
        \setcounter{figure}{0}
        \setcounter{table}{0}
        \setcounter{lemma}{0}
        \setcounter{theorem}{0}
        \setcounter{corollary}{0}
        \setcounter{definition}{0}
        \setcounter{equation}{0}
        \renewcommand{\thefigure}{\Alph{appendixc}.\arabic{figure}}
        \renewcommand{\thetable}{\Alph{appendixc}.\arabic{table}}
        \renewcommand{\theappendixc}{\Alph{appendixc}}
        \renewcommand{\thelemma}{\Alph{appendixc}.\arabic{lemma}}
        \renewcommand{\thetheorem}{\Alph{appendixc}.\arabic{theorem}}
        \renewcommand{\thedefinition}{\Alph{appendixc}.\arabic{definition}}
        \renewcommand{\thecorollary}{\Alph{appendixc}.\arabic{corollary}}
        \renewcommand{\theequation}{\Alph{appendixc}.\arabic{equation}}
        \noindent{\tenbf Appendix \theappendixc #1}\par\vspace{5pt}}
\newcommand{\subappendix}[1] {\vspace{12pt}
        \refstepcounter{subappendixc}
        \noindent{\bf Appendix \thesubappendixc. {\kern1pt \bfit #1}}
        \par\vspace{5pt}}
\newcommand{\subsubappendix}[1] {\vspace{12pt}
        \refstepcounter{subsubappendixc}
        \noindent{\rm Appendix \thesubsubappendixc. {\kern1pt \tenit #1}}
        \par\vspace{5pt}}

\newcommand{\smalllineskip}{\baselineskip=10pt}

\def\eightcirc{
\begin{picture}(0,0)
\put(4.4,1.8){\circle{6.5}}
\end{picture}}
\def\eightcopyright{\eightcirc\kern2.7pt\hbox{\eightrm c}}

\def\abstracts#1#2#3{{
        \centering{\begin{minipage}{4.5in}\baselineskip=10pt\footnotesize
        \parindent=0pt #1\par
        \parindent=15pt #2\par
        \parindent=15pt #3
        \end{minipage}}\par}}

\renewenvironment{thebibliography}[1]
        {\frenchspacing
         \ninerm\baselineskip=11pt
         \begin{list}{\arabic{enumi}.}
        {\usecounter{enumi}\setlength{\parsep}{0pt}
         \setlength{\leftmargin 12.7pt}{\rightmargin 0pt} 
         \setlength{\itemsep}{0pt} \settowidth
        {\labelwidth}{#1.}\sloppy}}{\end{list}}

\newcounter{itemlistc}
\newcounter{romanlistc}
\newcounter{alphlistc}
\newcounter{arabiclistc}

\newcommand{\fcaption}[1]{
        \refstepcounter{figure}
        \setbox\@tempboxa = \hbox{\footnotesize Fig.~\thefigure. #1}
        \ifdim \wd\@tempboxa > 5in
           {\begin{center}
        \parbox{5in}{\footnotesize\smalllineskip Fig.~\thefigure. #1}
            \end{center}}
        \else
             {\begin{center}
             {\footnotesize Fig.~\thefigure. #1}
              \end{center}}
        \fi}

\newcommand{\tcaption}[1]{
        \refstepcounter{table}
        \setbox\@tempboxa = \hbox{\footnotesize Table~\thetable. #1}
        \ifdim \wd\@tempboxa > 5in
           {\begin{center}
        \parbox{5in}{\footnotesize\smalllineskip Table~\thetable. #1}
            \end{center}}
        \else
             {\begin{center}
             {\footnotesize Table~\thetable. #1}
              \end{center}}
        \fi}

\def\@citex[#1]#2{\if@filesw\immediate\write\@auxout
        {\string\citation{#2}}\fi
\def\@citea{}\@cite{\@for\@citeb:=#2\do
        {\@citea\def\@citea{,}\@ifundefined
        {b@\@citeb}{{\bf ?}\@warning
        {Citation `\@citeb' on page \thepage \space undefined}}
        {\csname b@\@citeb\endcsname}}}{#1}}

\newif\if@cghi
\def\cite{\@cghitrue\@ifnextchar [{\@tempswatrue
        \@citex}{\@tempswafalse\@citex[]}}
\def\citelow{\@cghifalse\@ifnextchar [{\@tempswatrue
        \@citex}{\@tempswafalse\@citex[]}}
\def\@cite#1#2{{$\null^{#1}$\if@tempswa\typeout
        {IJCGA warning: optional citation argument
        ignored: `#2'} \fi}}

\def\@refcitex[#1]#2{\if@filesw\immediate\write\@auxout
        {\string\citation{#2}}\fi
\def\@citea{}\@refcite{\@for\@citeb:=#2\do
        {\@citea\def\@citea{, }\@ifundefined
        {b@\@citeb}{{\bf ?}\@warning
        {Citation `\@citeb' on page \thepage \space undefined}}
        \hbox{\csname b@\@citeb\endcsname}}}{#1}}

\def\@refcite#1#2{{#1\if@tempswa\typeout
        {IJCGA warning: optional citation argument
        ignored: `#2'} \fi}}

\def\refcite{\@ifnextchar[{\@tempswatrue
        \@refcitex}{\@tempswafalse\@refcitex[]}}

\def\pmb#1{\setbox0=\hbox{#1}
        \kern-.025em\copy0\kern-\wd0
        \kern.05em\copy0\kern-\wd0
        \kern-.025em\raise.0433em\box0}

\def\fnm#1{$^{\mbox{\scriptsize #1}}$}
\def\fnt#1#2{\footnotetext{\kern-.3em
        {$^{\mbox{\scriptsize #1}}$}{#2}}}

\def\fpage#1{\begingroup
\voffset=.3in
\thispagestyle{empty}\begin{table}[b]\centerline{\footnotesize #1}
        \end{table}\endgroup}

\def\runninghead#1#2{\pagestyle{myheadings}
\markboth{{\protect\footnotesize\it{\quad #1}}\hfill}
{\hfill{\protect\footnotesize\it{#2\quad}}}}
\font\tenrm=cmr10
\font\tenit=cmti10
\font\tenbf=cmbx10
\font\bfit=cmbxti10 at 10pt
\font\ninerm=cmr9

\font\eightrm=cmr8

\def\qed{\hbox{${\vcenter{\vbox{                      
   \hrule height 0.4pt\hbox{\vrule width 0.4pt height 6pt
   \kern5pt\vrule width 0.4pt}\hrule height 0.4pt}}}$}}

\begin{document}

\runninghead{A. E. Chubykalo \& S. J. Vlaev}
{Reply to ``Criticism of Necessity of simultaneous co-existence$\ldots$}

\thispagestyle{empty}\setcounter{page}{1}
\vspace*{0.88truein}
\fpage{1}

\centerline{\bf REPLY TO ``CRITICISM OF `NECESSITY OF SIMULTANEOUS
CO-EXISTENCE} \vspace*{0.035truein} \centerline{\bf
OF INSTANTANEOUS AND RETARDED INTERACTIONS}
\vspace*{0.035truein} \centerline{\bf
IN CLASSICAL ELECTRODYNAMICS' " by J.D.Jackson}
\vspace*{0.035truein}

\vspace*{0.37truein}
\centerline{\footnotesize ANDREW E. CHUBYKALO\fnm{*}\fnt{*}{E-mail:
andrew@logicnet.com.mx} AND STOYAN J. VLAEV}

\centerline{\footnotesize \it
Escuela de F\'{\i}sica, Universidad Aut\'onoma de Zacatecas, Apartado
Postal C-580}
\baselineskip=10pt
\centerline{\footnotesize
\it Zacatecas 98068, ZAC., M\'exico}


\baselineskip 5mm

\vspace*{0.21truein}

\abstracts{
In this note we show that Jackson's criticism of our work
``Necessity of simultaneous co-existence of\ldots" is based on an inexact
understanding of the basic assumptions and conclusions of our work.}{}{}


\bigskip

$$$$

In his note$^1$  J.D.Jackson  affirms that in our work$^2$ we
``{\sl make the claim that the electric and magnetic fields derived from
the Li\'enard-Wiechert potentials for a charged particle in arbitrary
motion do not satisfy the Maxwell equations}". This affirmation of 
J.D.Jackson does not correspond to a keynote of our work.

Let us begin with our general objections to  Jackson's criticism.
Actually, one of the aims of our work  was to show that the {\it direct}
use (from the mathematical point of view) of the following idea of Landau
and Lifshitz$^3$ (see the quotation below) leads to a contradiction:

\begin{quotation}
``{\sl To calculate the intensities of the electric and
magnetic fields from the formulas
\begin{equation}
{\bf E}=-\nabla\varphi-\frac{1}{c}\frac{\partial{\bf A}}{\partial
t},\qquad {\bf B}=[\nabla\times{\bf A}].
\end{equation}
we must differentiate $\varphi$ and {\bf A} with respect to the
coordinates x, y, z of the point, and the time t of observation. But the
formulas (63.5, {\rm Ref.3})

\begin{equation}
\varphi({\bf r},t)=\left\{\frac{q}{\left(R-{\bf R}\frac{{\bf
   V}}{c}\right)}\right\}_{t_0},\qquad
{\bf A}({\bf r},t)=\left\{\frac{q{\bf V}}{c\left(R-{\bf R}\frac{{\bf
   V}}{c}\right)}\right\}_{t_0}.
\end{equation}
express the potentials as a functions of $t_0$
($t^{\prime}$ {\rm in Ref.3})}, {\bf and only through} {\sl the relation
(63.1) {\rm in Ref.3)}}

 \begin{equation}
 t_0=t-\tau=t-\frac{R(t_0)}{c}.
 \end{equation}
{\bf as implicit} {\sl functions of $x,y,z,t$.
Therefore to calculate the required derivatives we must first calculate
the derivatives of $t_0$}"
\end{quotation}
In other words, if one takes into account exclusively the {\it implicit}
dependence of the potentials and fields on time $t$, one obtains correct
fields, but these fields

 \begin{equation}
   {\bf E}({\bf r},t)=q\left\{\frac{({\bf R}-R\frac{{\bf
         V}}{c})(1-\frac{V^{2}}{c^{2}})}{(R-{\bf R}\frac{{\bf
   V}}{c})^{3}}\right\}_{t_0}+q\left\{\frac{[{\bf R}\times[({\bf
   R}-R\frac{{\bf V}}{c})\times\frac{{\bf{\dot{V}}}}{c^{2}}]]}{(R-{\bf
   R}\frac{{\bf V}}{c})^{3}}\right\}_{t_0},
   \end{equation}
\begin{equation}
{\bf B}({\bf r},t)=\left\{\left[\frac{{\bf R}}{R}\times{\bf
E}\right]\right\}_{t_0},
\end{equation}
do not satisfy the Maxwell equations. {\bf Once more}: if, following
Landau and Lifshitz aforementioned idea, one does not take into account
the {\it explicit} dependence of fields on $t$, rather only the {\it
implicit} one, we can see that in this case (exclusively in {\it this}
case!) fields (4) and (5) do not satisfy Maxwell equations.

In  Section 4 of our work$^2$ we showed that {\it Faraday's law is
obeyed} if one considers the functions {\bf E} and {\bf B} as functions
with both {\it implicit} and {\it explicit} dependence on $t$ (or on
$x_i$).\fnm{a}\fnt{a}{By the way, Landau and Lifshitz in Ref.3 (we show
this below) do the same, conflicting with their phrase cited above.}
That is why we do not understand  why  J.D.Jackson did the same in
Section 4 of his work$^1$. It seems to us that a basic reason behind
Jackson's  antagonism to our work is the following: {\it our
interpretation of the explicit time-dependence as a certain manifestation
of instantaneous action-at-a-distance and on the other hand the implicit
time-dependence (i.e. exclusively through the relation (3)) as a
well-known short-range action}. From the {\it generally
accepted}\fnm{b}\fnt{b}{in works$^{4,5}$ we show that the generally
accepted point of view on the total and partial differentiation has some
serious problems} formal mathematical point of view our work is faultless.
Let us explain this point  more particularly.

In his work$^1$  J.D.Jackson considers our expression

\begin{equation}
\frac{\partial R}{\partial t_0}=-c
\end{equation}
as wrong, refering to formula ${\bf R}={\bf r}-{\bf r}_0(t_0)$. The point
is  what one means by the operator $\frac{\partial}{\partial t}$ and
by a function $R$ in Ref. 3. It is easy to prove that in the unnumerated
set of equations before  Eq.(63.6)$^3$ \begin{equation} \frac{\partial
R}{\partial t}= \frac{\partial R}{\partial t_0}\frac{\partial
t_0}{\partial t}= -\frac{{\bf RV}}{R}\frac{\partial t_0}{\partial t}=
c\left(1-\frac{\partial t_0}{\partial t}\right)
\end{equation}

Landau and Lifshitz mean by $\frac{\partial
R}{\partial t}$ the {\it total derivative} and {\it not partial one}!
(Recall that our index (0) corresponds to the index
($^{\prime}$) in Ref. 3).
In order to obtain a value of $\frac{\partial t_0}{\partial t}$ one cannot
perform the usual operation of differentiation. It is possible to calculate
this derivative  using a certain mathematical trick only. The authors of
REf. 3 use the fact that two {\it different} expressions of the  function
$R$ exist:
\begin{equation} R=c(t-t_0),\qquad {\rm where}\qquad
t_0=f(x,y,z,t), \end{equation} and \begin{equation}
R=[(x-x_0)^2+(y-y_0)^2+(z-z_0)^2]^{1/2},\qquad{\rm where}\qquad
x_{0i}=f_i(t_0).
\end{equation}

It is well-known from the classical analysis  that if a {\it given}
function is expressed by two different types of functional dependencies,
then exclusively {\it total} derivatives of these expressions with respect
to a {\it given} variable can be equated (contrary to the {\it partial}
ones). Comparing the total derivatives of $R$ from Eq.(8) and $R$ from
Eq.(9) Landau and Lifshitz obtain the corrected value of $\frac{\partial
t_0}{\partial t}$. So one can see that the expression $\frac{\partial
R}{\partial t}$ is the {\it total derivative} $\left(\frac
{dR}{dt}\right)$. And from

\begin{equation}
\frac{dR}{dt}= \frac{d}{dt}[c(t-t_0)]= \frac{\partial R}{\partial t}
\frac{\partial t}{\partial t}+\frac{\partial R}{\partial t_0}
\frac{\partial t_0}{\partial t}=
c\left(1-\frac{\partial t_0}{\partial t}\right)
\end{equation}
one can see that $\frac{\partial R}{\partial t}=c$ and
$\frac{\partial R}{\partial t_0}=-c$. However, one must not forget that
these expressions are just formal mathematical equalities and they do not
have any physical sense but help us to find a value of $\frac{\partial
t_0}{\partial t}$. Let adduce the
scheme which was implicitly used in Ref. 3 to
 obtain $\partial t_0/\partial t$ and $\partial t_0/\partial x_i$:

\bigskip
\bigskip
\clearpage

$$
\left[
\begin{array}{ccccc}
\underbrace{\frac{\partial R}{\partial t}_{(=c)}
+\frac{\partial R}{\partial t_0}_{(=-c)}
\frac{\partial t_0}{\partial t}}&=&
\underbrace{\frac{dR}{dt}}&=&
\underbrace{\sum\limits_k\frac{\partial R}{\partial x_{0k}}\frac{\partial
x_{0k}}{\partial t_0} \frac{\partial
t_0}{\partial t}}\\
\uparrow& &\uparrow& &\uparrow\\ R\{t,t_0(x_i,t)\}
&=&R(t_0)&=& R\{x_i,x_{0i}[t_0(x_i,t)]\}\\ \Updownarrow& &\Updownarrow&
&\Updownarrow\\
c(t-t_0)&=&R(t_0)&=&\left\{\sum_{i}[(x_i-x_{0i}(t_0)]^2\right\}^{1/2}\\
\downarrow& &\downarrow& &\downarrow\\
\overbrace{\frac{\partial R}{\partial t_0}_{(=-c)}
\frac{\partial t_0}{\partial x_i}}&=&
\overbrace{\frac{dR}{dx_i}}&=&
\overbrace{\frac{\partial R}{\partial
x_i}_{(=\frac{x_i-x_{0i}}{R})} + \sum\limits_k\frac{\partial R}{\partial
x_{0k}}\frac{\partial x_{0k}}{\partial t_0}
\frac{\partial t_0}{\partial x_i}} \end{array}\right].
$$
\begin{equation}
\end{equation}

\bigskip
\bigskip

If one takes into account that $\partial t/\partial x_i=\partial
x_i/\partial t=0$, as a result one obtains the correct expressions for
$\partial t_0/\partial t$ and $\partial t_0/\partial x_i$.

Finally, regarding two phrases of  J.D.Jackson in the
Abstract and at the close of  Ref. 1:
``{\sl Classical electromagnetic theory is complete as usually expressed}"
and ``{\sl Electromagnetic theory is complete in any chosen gauge}",
two sufficiently authoritative physicists of 20-th century help us:

R. Feynman$^6$:
\begin{quotation}
``...this tremendous
edifice (classical electrodynamics), which is such a beautiful success in
explaining so many phenomena, ultimately falls on its face.
...Classical    mechanics   is    a mathematically  consistent theory; it
just  doesn't agree  with experience. It is interesting, though, that the
classical  theory of electromagnetism is an unsatisfactory theory all by
itself.  There are  difficulties associated with the  ideas of Maxwell's
theory which are not solved by and not directly associated with quantum
mechanics..." \end{quotation}

W. Pauli$^7$:
\begin{quotation}
``We therefore see that {\it the Maxwell-Lorentz electrodynamics is
quite incompatible with the existence of charges, unless it is
supplemented by extraneous theoretical concepts}" (The choice
of italics was Pauli's).
\end{quotation}

\nonumsection{Acknowledgments}

We are grateful to  Annamaria D'Amore for revising the manuscript.

\nonumsection{References}

\end{document}